\documentclass[aps,prl,twocolumn,superscriptaddress]{revtex4-2}

\usepackage{enumitem}
\usepackage{graphicx}
\usepackage{amssymb}
\usepackage{amsmath}
\usepackage{bm}
\usepackage{xcolor}
\usepackage{braket}
\usepackage{kotex}
\usepackage[normalem]{ulem}

\usepackage{xfrac}
\usepackage{dsfont}
\usepackage{float}
\usepackage{dcolumn} 
\usepackage{mathtools}
\usepackage{hyperref}
\hypersetup{
     colorlinks=true,
     linkcolor=blue,
     filecolor=blue,
     citecolor=black,
     urlcolor=black,
     }
\usepackage[all]{hypcap}
\usepackage{enumerate}   
\usepackage{ulem}

\begin{document}

\title{Enhancement of damping in a turbulent atomic Bose-Einstein condensate}

\author{Junghoon Lee}
\affiliation{Department of Physics and Astronomy, Seoul National University, Seoul 08826, Korea}

\author{Jongmin Kim}
\affiliation{Department of Physics and Astronomy, Seoul National University, Seoul 08826, Korea}

\author{Jongheum Jung}
\affiliation{Department of Physics and Astronomy, Seoul National University, Seoul 08826, Korea}

\author{Y. Shin}
\email{yishin@snu.ac.kr}
\affiliation{Department of Physics and Astronomy, Seoul National University, Seoul 08826, Korea}
\affiliation{Institute of Applied Physics, Seoul National University, Seoul 08826, Korea}


\begin{abstract}
Turbulence enhances momentum transport in classical fluids, effectively increasing their viscosity. 
We investigate an analogous effect in a superfluid by measuring the damping of collective oscillations in an atomic Bose-Einstein condensate (BEC) containing stationary spin-superflow turbulence.
Using continuous spin driving to maintain turbulence in a spin-1 $^{23}$Na BEC, we excite its quadrupole mode and measure the damping rate over a range of temperatures. 
The damping consistently exceeds the Landau-damping rate expected for an equilibrium, non-turbulent BEC. 
The enhancement likely originates from two complementary processes: direct energy transfer from the mode to turbulent condensate fluctuations and turbulence-induced modification of the thermal cloud that amplifies Landau damping.
These results establish collective-mode damping as a sensitive probe of momentum transport in superfluid turbulence.
\end{abstract}

\maketitle

Turbulent or eddy viscosity is a phenomenological concept used to model the effects of turbulence on momentum transport within a fluid~\cite{pope2000turbulent,schmitt2007about}. 
This form of viscosity differs from the molecular viscosity in a laminar flow and originates from the macroscopic turbulent eddies that prevail in chaotic fluid motions.
The eddies interact dynamically across a wide range of length scales, thereby facilitating enhanced momentum exchanges and mixing within the fluid. 
Although turbulent viscosity, being an isotropic scalar quantity, appears too simplistic to capture the intricate dynamics of turbulent flows, it has nevertheless served as an effective and remarkably useful framework for predicting complex flow behaviors in various practical applications, from aerospace engineering \cite{leschziner2002turbulence,bhide2021improved} to ventilation in built environments~\cite{chen1995comparison}.

An intriguing question is whether the notion of turbulent viscosity can be extended to superfluids, which, due to their quantum nature, exhibit zero intrinsic viscosity~\cite{vinen2002quantum,barenghi2014experimental}.
Despite their inviscid character, superfluids can still dissipate kinetic energy through the dynamics of quantum vortices~\cite{pu1999coherent,bewley2008characterization} and their interactions with the coexisting normal fluid component~\cite{hall1956rotation,skrbek2021phenomenology}.
Notably, energy cascades and Kolmogorov-like scaling behaviors have been observed in turbulent superfluids~\cite{maurer1998local,navon2016emergence,zhao2025kolmogorov}, pointing to a striking resemblance with classical turbulence.
Early studies of superfluid helium turbulence introduced the concept of eddy viscosity to account for excess pressure gradients observed under heat counterflow conditions~\cite{brewer1961heat,eddyviscosityHe4}, but the interpretation remained inconclusive due to uncertainties in turbulence homogeneity and ambiguities regarding the state of the normal fluid~\cite{kashiwamura1999effects}.
More recently, piston-driven shock experiments in atomic Bose-Einstein condensates (BECs) revisited the eddy viscosity framework to describe dissipative shock dynamics~\cite{mossman2018dissipative}.
A coherent understanding of turbulence-induced viscous effects in superfluids is therefore highly desirable, with potential implications for astrophysical contexts such as neutron stars, which are believed to host superfluid interiors~\cite{melatos2014pulsar,andersson2007superfluid}.

\begin{figure}[t]
\includegraphics[width=8.6cm]{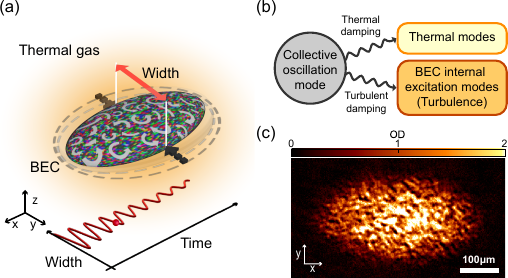}
\centering
\caption{Collective oscillations of a turbulent Bose-Einstein condensate (BEC). (a) Schematic of the experiment. A BEC with internal turbulent flow (gray arrows) is confined in a harmonic potential and undergoes shape oscillations. The turbulence is sustained by resonant RF spin driving, which steadily generates an irregular spin texture (color pattern).  (b) Energy flow diagram illustrating two pathways for dissipation of collective excitation energy: direct interaction with thermal components and energy transfer into the internal turbulence. (c) Image of a turbulent BEC after an 18-ms time-of-flight.}
\label{fig1}
\end{figure}

In this Letter, we experimentally investigate turbulence-enhanced momentum transport in atomic BECs. 
Using continuous spin driving, we create a spin-1 BEC in a non-equilibrium steady state that hosts turbulent spin-superflow~\cite{hong2023spin,jung2023random,fujimoto2014}. We then examine how this turbulence affects the damping of collective quadrupole oscillations of the BEC [Fig.~1(a)].
By comparing systems with and without turbulence, and measuring the temperature dependence of the damping rates, we find that turbulent BECs exhibit enhanced damping relative to the Landau-damping expectation for equilibrium condensates~\cite{pethick2002bose,fedichev1998damping}.
We express this excess in terms of an effective kinematic viscosity, $\nu_\text{T}$, directly analogous to the turbulent viscosity of classical fluids~\cite{landau1987fluid,schmitt2007about}, and discuss plausible pathways for damping enhancement.

Our superfluid system consists of a BEC of $^{23}$Na atoms in the $F$=1 hyperfine state, which has internal spin degrees of freedom. 
We prepare a BEC initially in the $m_F=-1$ spin state within an optical dipole trap (ODT) under a uniform external magnetic field. Turbulence is generated  using a spin-driving technique as described in~\cite{hong2023spin}, in which a radio-frequency (RF) magnetic field is applied at the Larmor resonance frequency.
Under resonant spin driving the spin dynamics becomes chaotic~\cite{rautenberg2020classical,evrard2021many}, giving rise to an irregular spin texture across the BEC~\cite{kim2024chaos}. With continuous driving, the system evolves to a non-equilibrium steady state with equal populations of the three, $m_F=\pm1, 0$ spin components, where the spin texture is persistently randomized and the associated superflow turbulence is sustained over time~\cite{hong2023spin,kim2024chaos}. Figure~\ref{fig1}(c) shows a time-of-flight image of such a turbulent BEC, where turbulence is evidenced by irregular density modulations.
The heating rate from the resonant spin driving is negligible, allowing a turbulent BEC to maintain a long lifetime comparable to the vacuum-limited lifetime ($>30$~s)~\cite{hong2023spin}. This gentle turbulence generation scheme enables the investigation of collective oscillations in a BEC sustained in a steady turbulent state.

Collective oscillations of a trapped BEC typically exhibit damping in the presence of a coexisting thermal cloud. This thermal dissipation is well described by the Landau damping mechanism, in which the energy and momentum of collective excitations are transferred to thermally excited particles~\cite{landaudamping,pethick2002bose}.
For low-lying collective excitations, the damping rate $\Gamma_{th}$  at temperatures $k_\text{B} T > \mu$ is given by~\cite{fedichev1998damping} 
\begin{eqnarray}
\label{Landaudamp}
 \frac{\Gamma_{th}}{\omega_\nu} = A_\nu \sqrt{n_0 a^3}~ \frac{k_\text{B} T}{\mu},
 \label{eq:Fedichev}
\end{eqnarray}
where $\omega_\nu$ is the oscillation frequency, $n_0$ is the peak condensate density, $a$ is the scattering length, $k_\text{B}$ is the Boltzmann constant, and $\mu$ is the chemical potential. 
The factor $A_\nu$ is a dimensionless parameter determined by the trap geometry and the characteristics of the collective mode $\nu$~\cite{chevy2002breathing,jackson2003landau}.
This form of thermal damping has been extensively demonstrated in previous experiments for various collective modes of trapped BECs~\cite{jin1997temperature,stamper1998collisionless,jackson2002quadrupole,meppelink2009damping,yuen2015enhanced, morgan2003quantitative}.

In a turbulent BEC, nonlinear coupling between collective modes can provide an extra damping channel by diffusing long-wavelength momentum throughout the turbulent medium.
Monitoring how turbulence alters the damping behavior of low-lying collective excitations therefore provides a sensitive probe of turbulence-induced dissipation.
This approach is particularly effective because it relies on well-established collective modes that have been widely used to characterize finite-temperature effects in BECs.

\begin{figure}[t]
\includegraphics[width=8.6cm]{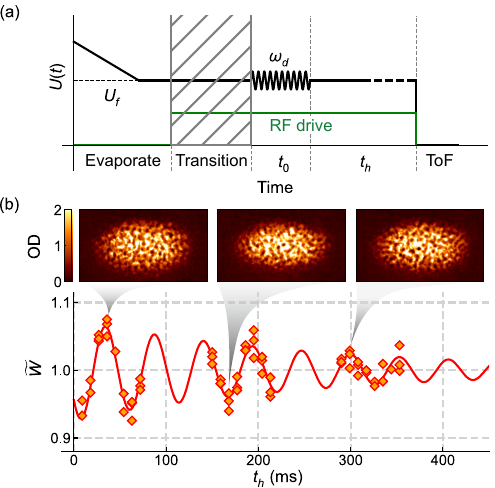}
\centering
\caption{Observation of damped oscillations of a turbulent BEC. (a) Experimental sequence. The sample temperature is controlled by adjusting the final trap depth $U_f$ during evaporation cooling. After cooling, sample is held for a few seconds to thermalize and damp out residual motion. A resonant RF magnetic field is represented as a green solid line, continuously applied to generate and sustain turbulence. The hatched area indicates the period of transition to steady turbulence. After a steady turbulent state is prepared, the trap is perturbatively modulated during a time $t_0$, to excite the quadrupole mode. After a variable hold time $t_h$, an absorption image is taken after time-of-flight of 18-ms. (b) Time evolution of the normalized condensate width $\widetilde{W}$ along the $y$ direction as a function of the hold time $t_h$. Images at top were taken at $t_h=36$-ms, $168$-ms, and $299$-ms, from left to right, respectively. Each data point represents a single measurement and the solid red line is a damped sinusoidal function fit to the mean values of the data. This fit yields an oscillation frequency $\omega_\nu=18.8(9)$ Hz and damping rate $\Gamma=0.60(9)$ Hz. The thermal fraction of the sample was $0.44(1)$.}
\label{fig2}
\end{figure}

In Fig.~2(a), we delineate our experimental procedure to investigate the damping of collective oscillations. After preparing a BEC in a steady turbulent state by applying a RF field, we drive collective excitations by modulating the power of the ODT laser beam. The depth of the optical trap is modulated as $U(t)=U_{f}[1+\epsilon \sin (\omega_d t)]$ for a short period set to $t_0=10\frac{2\pi}{\omega_d}$, with its frequency $\omega_{d}$ and relative amplitude factor $\epsilon$. 
For our highly oblate sample with trapping frequency ratios of $\omega_x: \omega_y: \omega_z \approx 1:2:100$, two distinct quadrupole modes are identified with oscillation frequencies of $\omega_\nu\approx 0.8\,\omega_y$ and $1.6\,\omega_y$~\cite{SM}. The high (low) frequency mode exhibits in-phase (out-of-phase) oscillations in condensate widths along the $x$ and $y$ directions~\cite{stringari1996collective,Heiselberg2004collective}. In the present work, we investigate the high-frequency mode in which the BEC experiences a higher shear flow during oscillations, offering a favorable setting for exploring turbulent viscosity~\cite{SM,cao2011,vogt2012scale}.
To selectively excite the quadrupole mode and minimize interference from other modes, $\omega_d$ was tuned to the red side of the resonance, further away from the closest thermal gas mode at $2\omega_y$~\cite{jin1997temperature,stamper1998collisionless,meppelink2009damping,SM}.

The oscillations of the BEC are measured by tracking the normalized width $\widetilde{W}(t_h)$ along the $y$ direction after a variable hold time $t_h$ and subsequent time-of-flight imaging [Fig.~\ref{fig2}(b)]. 
Here, $\widetilde{W}=W/W_\text{eq}$ , with $W$ being the measured width and $W_\text{eq}$ the width of an unmodulated BEC for the measured atom number $N_c$ of the condensate~\cite{SM}.
The oscillating behavior is well described by an exponentially damped sinusoid, $\widetilde{W}(t_h)=1+ A e^{-\Gamma t_h} \sin (\omega_\nu t_h +\phi)$, where $A$ and $\phi$ are the relative amplitude and phase of the oscillation, respectively.
The persistence of a well-defined mode indicates that the hydrodynamic nature is preserved in the turbulent BEC, ensuring that the mode's damping rate offers a quantitative probe of the effect of the turbulence.
In our measurements, we keep the in-trap oscillation amplitude below 10\%~\cite{SM}.

To demonstrate damping enhancement due to turbulence, we first need an appropriate reference for comparison.
The most natural baseline is a thermal equilibrium sample that contains the same numbers of thermal and condensed atoms, $N_{th}$ and $N_c$, as the turbulent sample and also has the same spin composition.
In the steady turbulent state under continuous spin driving, the thermal cloud is an equal mixture of three spin components~\cite{hong2023spin}, and the corresponding equilibrium temperature is $k_\text{B} T= 0.94 \hbar\overline{\omega}({N_{th}}/\mathcal{D}_{s})^{1/3}$ 
with $\bar{\omega} = (\omega_x \omega_y \omega_z)^{1/3}$ and  $\mathcal{D}_{s}$=3 denoting the number of spin components~\cite{pethick2002bose}.
However, a difficulty arises from the fact that a spin-1 BEC at thermal equilibrium tends to develop magnetic ordering due to spin interactions~\cite{kawaguchi2012spinor}, which disfavors a perfectly spin-symmetric thermal cloud. Therefore, directly measuring the reference damping rate for a $\mathcal{D}_{s}$=3 thermal-equilibrium sample is impractical. Nevertheless, since the number of thermal atoms scales with $\mathcal{D}_{s}$ while their momentum distribution at temperature $T$ remains unchanged, the damping rate should increase by the same factor, provided the scattering properties are identical for all spin components~\cite{kawaguchi2012spinor, bhattacherjee2014damping}.
Accordingly, we estimate the reference damping rate from Eq.~\eqref{Landaudamp} using $A_\nu = \mathcal{D}_{s} A_\nu^{(0)}$, where $A_\nu^{(0)}$ is the value of a single component system at thermal equilibrium.

With this reference damping rate in hand, we investigate the temperature dependence of the damping rates for both spin-driven turbulent BECs ($\mathcal{D}_{s}=3$) and single-component BECs ($\mathcal{D}_{s}=1$) prepared without RF spin driving. The temperature is
controlled by adjusting the final trap depth $U_f$ during evaporation cooling [Fig.~2(a)]. As $U_f$ increases, the final trapping frequencies range from $2\pi \times (4.4, 8.9, 420)$ Hz to $2\pi \times (7.9, 16.3, 780)$ Hz, and correspondingly the thermal fraction changes approximately from 0.25 to 0.7 with turbulence and from 0.1 to 0.5 without turbulence.

\begin{figure}[t]
\includegraphics[width=8.6cm]{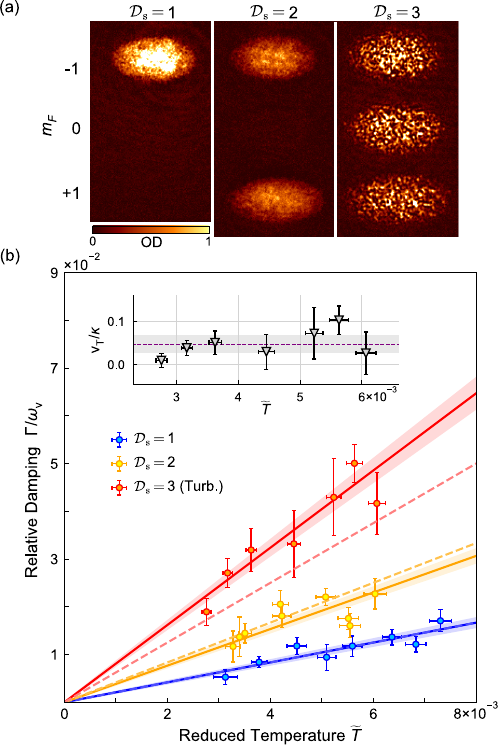}
\centering
\caption{\label{fig3} 
Temperature dependence of the damping of collective oscillations of BECs.
(a) Images of the three different types of BEC samples under investigation: single-component ($\mathcal{D}_{s}=1$) and two-component ($\mathcal{D}_{s}=2$) BECs without turbulence, and a three-component turbulent BEC ($\mathcal{D}_{s}=3$). To visualize the spin composition, the images were taken after Stern-Gerlach spin separation during free expansion~\cite{footnote_SG}.
(b) Relative damping rates $\Gamma/\omega_\nu$ as functions of the reduced temperature $\widetilde{T}$ for $\mathcal{D}_{s}=1,2,3$. The solid lines represent weighted linear fits of $\Gamma/\omega_\nu=A_\nu \widetilde{T}$ [Eq.~(1)] to the data, and the shaded regions indicate the 1$\sigma$ uncertainties of the fits including the individual uncertainties of the data points. The dashed lines show the predictions of the thermal damping model for $\mathcal{D}_{s}=2,3$, respectively, based on the measurement data for $\mathcal{D}_{s}=1$.
The inset shows the effective kinematic viscosity $\nu_\text{T}$ calculated from the excess damping $\Gamma_\text{T}$ in the turbulent BECs [Eq.~(2)]~\cite{SM} with $\kappa=h/m$. The horizontal dashed line marks the average of the $\nu_\text{T}$ values and the shaded area indicates its $1\sigma$ standard error.}
\end{figure}

Furthermore, to verify the dependence of $A_\nu$ on $\mathcal{D}_{s}$, we conduct a parallel experiment with thermal equilibrium samples having two spin components ($\mathcal{D}_{s}=2$) [Fig.~3(a)], which are equal mixtures of the $m_F=\pm 1$ components that are miscible for our $^{23}$Na BEC system~\cite{stenger1998spin}. 
To prepare them, the atoms are transferred to the $m_F=0$ state using a rapid adiabatic passage with an RF magnetic field, followed by a pulse of a strong magnetic field gradient to purify the spin state. We then applied a $\pi/2$ resonant RF magnetic pulse to form an equal mixture of the $m_F=\pm 1$ components. 
After the preparation of the spin mixture, the quadratic Zeeman energy is changed from positive to small negative value to prevent spontaneous formation of the $m_F=0$ population and stabilize the two-component system \cite{gerbier2006resonant,seo2015half}.
In all the damping-rate measurements, we nulled the magnetic field gradient to below 0.1~mG/cm to suppress spin-drag effects in the collective oscillations~\cite{duine2009spin}. The residual field gradient was calibrated and minimized using a Ramsey interferometry technique~\cite{jpsj2013,hong2023spin,seo2015half}.

Figure~\ref{fig3}(b) shows the measurement results of the relative damping rates $\Gamma/\omega_\nu$ for samples with one, two and three spin components ($\mathcal{D}_{s}=1,2,3$) as functions of the reduced temperature $\widetilde{T}\equiv k_\text{B}T \sqrt{n_0a^3}/\mu$.
The scattering length $a$ varies by up to 7\% among the spin components~\cite{knoop2011}, but because $\widetilde{T}\propto a^{4/5}$~\cite{footnote2}, this variation alters $\widetilde{T}$ by $\lesssim 6$\%. For simplicity, we use the $m_F=-1$ value of $a$ for all data sets~\cite{footnote_scat}. 
For $\mathcal{D}_s=1,2$, we extract the Landau-damping prefactor $A_\nu$ by linear fits to Eq.~(\ref{eq:Fedichev}), obtaining $A_\nu=2.1(1)$ and $3.8(2)$, respectively.
For the turbulent $\mathcal{D}_{s}=3$ sample, we do not assume the applicability of Eq.~(\ref{eq:Fedichev}); instead, we use $\widetilde{T}$ only as a scaling variable to facilitate a quantitative comparison with an equilibrated reference, yielding an effective slope $A_\nu=8.1(4)$.

We note that the single-component value, $A_\nu^{(0)}\equiv A_\nu(\mathcal{D}_{s}=1)$, is smaller than typical values reported for cylindrically symmetric traps~\cite{jin1997temperature}. Such variations are expected because $A_\nu$ depends sensitively on trap geometry and mode structure in a trapped gas~\cite{fedichev1998damping,chevy2002breathing,jackson2003landau}. We attribute the reduced $A_\nu^{(0)}$ to the highly oblate trap geometry and tight confinement along $z$, which may place the system in a dimensional-crossover regime~\cite{yuen2015enhanced,footnote3}.
The two-component result satisfies $A_\nu=2A_\nu ^{(0)}$ within experimental uncertainty, confirming that the thermal damping rate scales linearly with the number of spin components.
We therefore use the measured $A_\nu^{(0)}$ as an empirical baseline for our specific trap geometry and quadrupole mode.

Relative to this calibrated benchmark, the turbulent three-component sample yields $A_\nu>~3A_\nu^{(0)}$, indicating damping beyond the equilibrium expectation. We quantify the excess damping as 
\begin{equation}
    \Gamma_\text{T}=\Gamma-3 A_\nu^{(0)} \widetilde{T} \omega_\nu
\end{equation}
and attribute it to turbulence-induced damping. Averaged over the temperature range explored, we find $\Gamma_\text{T}\approx 2\pi \times 0.2$~Hz.

Following the phenomenological approach in classical fluid dynamics, we recast the excess damping $\Gamma_\text{T}$ in terms of turbulent viscosity $\nu_\text{T}$~\cite{landau1987fluid,kavoulakis1998}.
In classical turbulence, Reynolds-averaged Navier-Stokes (RANS) equations absorb the time-averaged influence of velocity fluctuations into a single parameter $\nu_\text{T}$, providing a closed model of mean flow~\cite{pope2000turbulent,schmitt2007about}.
Adopting this framework to our steady spin-superfluid turbulence and modeling the quadrupole oscillations with the mean velocity field of $\boldsymbol{v}(\boldsymbol{r},t) = (b_x x, b_y y, b_z z)\sin{\omega_\nu t}$, we relate the measured excess damping to a shear-stress action proportional to $\nu_\text{T}$~\cite{SM}. We obtain an effective kinematic viscosity of $\nu_\text{T} = 0.05(2) \kappa$, where $\kappa=h/m$ with $h$ being the Planck constant and $m$ the atomic mass. Interestingly, this value is of the same order as the effective viscosity $\sim 0.1 \kappa$ reported for turbulent superfluid $^4$He in the $T\rightarrow 0$ limit~\cite{walmsley2008quantum,zmeev2015dissipation,babuin2014effective}. However, direct comparison is tentative because the measurements stem from different experimental contexts and no unified theoretical framework yet connects them~\cite{shukla2019quantitative}.

The key question now concerns the physical origin of the enhanced damping.
Two mechanisms are most plausible. 
First, the quadrupole mode's shear stress can transfer energy directly to the turbulent condensate fluctuations, a purely hydrodynamic pathway that underlies our RANS-based estimation of turbulent viscosity.  
Second, turbulence can amplify Landau damping by reshaping the surrounding thermal cloud.
Because Landau damping of low-lying modes is highly sensitive to the occupation of thermal states near the chemical potential, even modest, non-equilibrium turbulence can redistribute those occupations and alter the damping rate. 
This feedback loop goes beyond the conventional view of the thermal component as merely a high-momentum energy sink via mutual friction with quantum vortices~\cite{amette2025vortex}.
Instead, it highlights the full two-fluid character of the superfluid system and may have important implications for wave turbulence in the atomic BEC system.
Disentangling the relative contributions of these two channels will require a detailed characterization of the spin-superflow turbulence, such as its energy spectrum, cascade dynamics, and reciprocal coupling to the thermal gas~\cite{fujimoto2014,jung2023random}.

Finally, we note that the turbulent flow in our system is predominantly two-dimensional due to the oblate trap, whereas the mean flow associated with the collective modes remains three-dimensional. This disparity may limit the use of the standard RANS framework, which presupposes three-dimensional homogeneous isotropic turbulence. 
Studies of classical turbulence driven by two-dimensional forcing have shown that introducing stronger three-dimensionality can increase the eddy viscosity by suppressing the inverse energy cascade~\cite{shats2010turbulence}.
Mapping how the effective turbulent viscosity evolves across the 2D-3D crossover therefore represents a promising avenue for future work.

In summary, by continuously driving spin dynamics to maintain a steady turbulent state, we have observed an enhancement of damping in the collective oscillations of a turbulent BEC.
Viewing this excess through the lens of an effective turbulent viscosity provides a new framework for probing energy-transfer pathways and hydrodynamic properties in superfluid turbulence.
This work can extend to other collective excitation modes, particularly including a persistent monopole (breathing) mode available in spherically symmetric BECs~\cite{lobser2015observation}.

\begin{acknowledgments}
This work was supported by the National Research Foundation of Korea (Grants No. RS-2023-NR077280, No. RS-2023-NR119928, and No. RS-2024-00413957).
\end{acknowledgments}

\clearpage
\onecolumngrid

\setcounter{equation}{0}
\setcounter{figure}{0}
\setcounter{table}{0}
\makeatletter
\renewcommand{\theequation}{S\arabic{equation}}
\renewcommand{\thefigure}{S\arabic{figure}}
\renewcommand*{\bibnumfmt}[1]{[S#1]}
\setcounter{secnumdepth}{3}
\setcounter{subsection}{0}
\renewcommand{\thesubsection}{\Alph{subsection}}
\renewcommand{\thesubsubsection}{\arabic{subsubsection}}
\makeatother

\begin{center}
  \large\bfseries Supplemental Material for\\
Enhancement of damping in a turbulent atomic Bose-Einstein condensate
\end{center}






\setcounter{equation}{0}
\setcounter{figure}{0}
\setcounter{table}{0}
\makeatletter
\renewcommand{\theequation}{S\arabic{equation}}
\renewcommand{\thefigure}{S\arabic{figure}}
\renewcommand*{\bibnumfmt}[1]{[S#1]}
\renewcommand\NAT@open{[S}
\renewcommand\NAT@close{]}
\makeatother

\subsection{Methods}
\subsubsection{Sample preparation}

We prepared a thermal cloud of $^{23}$Na atoms in the $|F=1,m_F=-1\rangle$ state in an optical dipole trap (ODT) and applied evaporative cooling by reducing the trap depth to produce a Bose-Einstein condensate (BEC). 
The thermal fraction $\zeta_{th}$ of the sample was controlled by adjusting the final value $U_f$ of the trap depth. 
At our lowest temperatures, the typical number of atoms for the BEC was approximately $7 \times 10^6$.
The atomic sample was held in the ODT for several seconds to relax residual motion before driving the collective excitations of the trapped BEC.

Depending on the target experiment, the atomic sample was transformed into a two-component system with $|m_F=\pm 1\rangle$ states ($\mathcal{D}_s=2$) or a turbulent system with $|m_F=0,\pm1\rangle$ states ($\mathcal{D}_s=3$). For the two-component system, we first transferred atoms to the $|m_F=0\rangle$ state using a rapid adiabatic passage with an RF magnetic field, followed by a pulse of a strong magnetic field gradient to purify the spin state. We then applied a $\pi/2$ resonant RF magnetic pulse to form an equal mixture of the $m_F=\pm 1$ components. After the preparation of the spin mixture, we irradiated a continuous microwave field detuned from the $F=1 \rightarrow F=2$ transition to shift the quadratic Zeeman energy to a small negative value. This prevents the creation of atoms in the $|F=1,m_F=0\rangle$ state via spin exchange processes, stabilizing the two-component system \cite{sgerbier2006resonant,sseo2015half}.

The turbulent sample was prepared with the continuous application of the resonant RF field~\cite{shong2023spin}. 
A steady turbulent state with an irregular spin texture was generated within 0.5~s, exhibiting equal populations of the three $m_F=0,\pm1$ spin components. This turbulence is sustained by chaotic spin dynamics under RF spin driving~\cite{skim2024chaos}, as detailed in Section B. Following 2~s of RF driving, we conducted the collective oscillation experiment while maintaining the RF spin driving.
The BEC lifetime exceeded 30~s under RF driving ($\mathcal{D}_s=3$) and was approximately 5~s under the microwave dressing ($\mathcal{D}_s=2$), significantly longer than our measurement duration, which lasted for a few hundreds of ms.

The ODT was highly oblate such that the ratio of the trapping frequencies was $\omega_x: \omega_y: \omega_z \approx 1: 2: 100$.
The trapping frequencies were measured by analyzing the dipole oscillations of a trapped BEC along the $x$ and $y$ directions and the parametric heating of trap modulations for the $z$ axis. 

\subsubsection{Imaging}
An absorption image of the sample was taken in the $z$ direction after a time of flight of $\tau=18$~ms by releasing the trapping potential. 
The number of atoms contained in the thermal gas, $N_{th}$, was estimated from a two-dimensional Gaussian distribution fit to the outer region of the cloud, and the number of condensed atoms $N_c$ was determined from an image obtained by subtracting the fitted thermal profile from the original image. The thermal fraction is given by $\zeta_{th}=N_{th}/N$ with $N=N_c+N_{th}$.
The width $W_\alpha(t)$ of the condensates along the $\alpha$ axis ($\alpha = x,y,z$) was determined by fitting the two-dimensional Thomas-Fermi profile to the subtracted image.

\subsubsection{Analysis of width oscillations}
\label{widthosc}
The width $W(t)$, measured after the free expansion for $\tau$, is related with the in-situ width $W_i(t)$ of the BEC as
\begin{equation}
W(t)=W_{i}(t)+\frac{dW_{i}(t)}{dt}\tau.
\label{tof_width}
\end{equation}
The second term on the right hand side accounts for the width change during the time of flight, owing to the oscillation velocity of the condensate.
Here we neglect the expansion effect due to the mean-field energy and quantum pressure, which are negligibly small along the $y$ direction for our highly oblate geometry. 
When a trapped BEC oscillates as $W_i(t)=W_\text{eq}(1+ B\sin(\omega t+\theta))$, using Eq.~(\ref{tof_width}), the measured width is given by $W(t)=W_\text{eq}(1+ \alpha_\tau B \sin(\omega t + \theta+\varphi_\tau) )$ with 
$\alpha_\tau = \sqrt{1 + (\omega \tau)^2}$ and $\varphi_\tau = \tan^{-1}(\omega \tau)$ account for the effects of amplitude amplification and the phase change, respectively, due to time-of-flight expansion.
$W_\text{eq}$ denotes the in-situ width of the condensate at equilibrium. 

In our data analysis, we used the normalized width $\widetilde{W}= W/W_\text{eq}$ to account for variations in the number of atoms in the sample. Using the relationship of $W_\text{eq}\propto U_f^{-1/5} N_c^{1/5}$ in the Thomas-Fermi (TF) approximation, the equilibrium width for a given $N_c$ was estimated as $W_\text{eq}(N_c)=\overline{W}_\text{eq} (N_c/ \overline{N_c})^{1/5}$, where $\overline{W}_\text{eq}$ and $\overline{N_c}$ are the mean width and the mean number of atoms of the condensate, as determined by averaging tens of measurements without trap modulations.

\subsection{Stationary spin-superflow turbulence}
Turbulence in a spin-1 BEC was induced by an RF magnetic field applied transversely to a uniform external field $B_z$.
In a mean-field description, neglecting the spatial modes of the BEC and taking the rotating wave approximation, the local dynamics of the spin state $\boldsymbol{\zeta}=(\zeta_{+1}, \zeta_{0},\zeta_{-1})^\text{T}$ of the BEC is governed by the following Hamiltonian per particle,
\begin{equation}
 H_s = \hbar \delta f_z  -\hbar \Omega f_x + q \boldsymbol{\zeta}^{\dagger}\text{f}_z^2\boldsymbol{\zeta} + \varepsilon_s \vert \boldsymbol{f} \vert^2,
 \label{Hamiltonian}
 \end{equation} 
where $\textbf{f} = (\text{f}_x,\text{f}_y,\text{f}_z)$ are the spin operators of the spin-1 system and $\boldsymbol{f} = \boldsymbol{\zeta}^{\dagger} \textbf{f} \boldsymbol{\zeta}$ is the normalized spin vector with $f_{x,y,z}$ representing the normalized magnetizations in $x,y,$ and $z$ directions, respectively. In addition, $\delta=\omega-\omega_0$ is the frequency detuning of the RF magnetic field from the Larmor frequency $\omega_0= \frac{1}{2} \mu_\text{B} B_z/\hbar$ with $\mu_\text{B}$ being the Bohr magneton, $\Omega$ is the Rabi frequency of the RF field, and $q$ denotes the quadratic Zeeman energy. The last term represents the spin interaction energy and $\varepsilon_s>0$ for antiferromagnetic interactions. In our experiment, $\Omega = 2\pi \times 150$~Hz, $q/h = 47$ Hz, and $\varepsilon_s/h$ ranges from 37~Hz to 80~Hz for the peak atom density.

When the system is driven resonantly ($\delta= 0$) and the energy scales of $\hbar\Omega$, $q$, and $\varepsilon_s$ are comparable, the spin dynamics of the Hamiltonian $H_s$ becomes chaotic~\cite{srautenberg2020classical,sevrard2021many}. 
Furthermore, in the experiment, the external magnetic field was slightly modulated due to field noises ($\sim 1$~mG), which was found to enhance the chaoticity of the system, thus facilitating complete randomization of the spin state~\cite{skim2024chaos,sjung2023random}.
Due to this chaotic spin dynamics, small spatial fluctuations in the wavefunction of the BEC, even when starting with a uniform spin texture, develop into complex spatial variations of the spin states, resulting in an irregular spin texture.

In a spinor BEC, a superflow is associated not only with the spatial variations of the superfluid phase $\varphi$ but also with those of $\boldsymbol{\zeta}$, i.e., the spin texture~\cite{skawaguchi2012spinor}. 
The superfluid velocity $\boldsymbol{v}$ and corresponding vorticity are given by
\begin{eqnarray}
    \boldsymbol{v}= \frac{\hbar}{m}(\boldsymbol{\nabla} \varphi - i\boldsymbol{\zeta}^{\dagger}\boldsymbol{\nabla}\boldsymbol{\zeta}), \nonumber \\
    \boldsymbol{\nabla} \times \boldsymbol{v} = -\frac{i\hbar}{m} \boldsymbol{\nabla} \boldsymbol{\zeta}^{\dagger} \times \boldsymbol{\nabla} \boldsymbol{\zeta},
    \label{superfluid}
\end{eqnarray}
where $m$ is the particle mass. Therefore, when the spin texture is continuously driven to be randomized due to chaotic spin dynamics, the associated turbulent flow in the BEC is sustained~\cite{skim2024chaos}. 

\subsection{Quadrupole modes of highly oblate condensates}
\label{Quadrupole mode calc}
In this section, we briefly describe the collective quadrupole excitation modes of a BEC trapped in a harmonic potential, following earlier works ~\cite{sstringari1996collective,sHeiselberg2004collective}.
Based on the time-dependent Gross-Pitaevskii equation for a macroscopic wavefunction of condensate $\psi(\boldsymbol{r},t) \equiv \sqrt{n(\boldsymbol{r},t) }e^{i\varphi(\boldsymbol{r},t)}$,
where $n$ is the local particle density of the condensate, the equations of motion for the BEC are given by
\begin{align}
    \frac{\partial n}{\partial t} + &\boldsymbol{\nabla} \cdot (n\boldsymbol{v})=0, \label{conti}
     \\
    \frac{\partial\boldsymbol{v}}{\partial t}+(\boldsymbol{v}\cdot\boldsymbol{\nabla})\boldsymbol{v}=-\frac{1}{m}&\boldsymbol{\nabla}\left(V_\text{ext}+g n-\frac{\hbar^2}{2m}\frac{ \boldsymbol{\nabla}^2\sqrt{n}}{\sqrt{n}}\right),
    \label{hydro}
\end{align}
where $\boldsymbol{v}= \frac{\hbar}{m}\boldsymbol{\nabla}\varphi$ is the superfluid velocity field, $g$ represents the interaction strength, and $V_\text{ext}(\boldsymbol{r}) = \frac{1}{2}m\sum_{\alpha} \omega_\alpha^2 r_\alpha^2$ is the external harmonic potential, with $r_\alpha$ representing the coordinates along the $\alpha$ axis ($\alpha = x,y,z$). 
Eq.~(\ref{conti}) is the continuity equation for the conservation of the particle number and Eq.~(\ref{hydro}) is referred to as the quantum Navier-Stokes equation (or the Euler equation for a superfluid).
Using the Thomas-Fermi (TF) approximation, the last term of Eq.~(\ref{hydro}), called the quantum pressure, is neglected and the equilibrium density profile of the BEC is given as $n_0(\bold{r})=\max[\frac{\mu-V_\text{ext}(\bold{r})}{g},0]$,
where $\mu$ is the chemical potential of the condensate. Note that this approximation is not reliable near the boundary region at a low particle density.

\begin{figure}[t]
    \includegraphics[width=16.1cm]{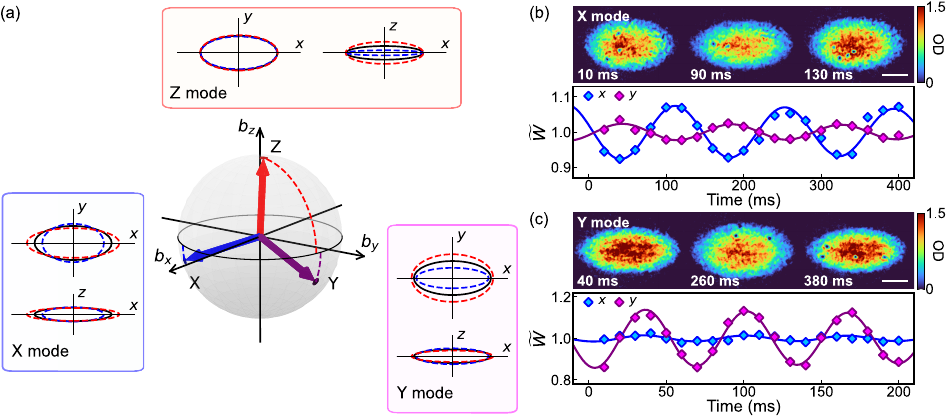}
    \centering 
\caption{(a) Quadrupole oscillation modes for an oblate BEC with trapping frequencies $\omega_x<\omega_y<\omega_z$. The oscillatory motion is characterized by $(b_x, b_y, b_z)$, as shown in Eq.~(\ref{velocity}), and the three modes are denoted by $\text{X, Y,}$ and $\text{Z}$, respectively, corresponding to the dominant oscillation axis. The variation in the density profile in the $xy$ and $xz$ planes for each mode is presented.
(b),(c) Experimental observation of the quadrupole oscillations of BECs. BEC samples with $\mathcal{D}_s=1$ were prepared at our lowest temperatures and the $\text{X}$ or $\text{Y}$ mode was selectively excited by short trap modulations at a frequency slightly red-detuned from resonance.  
The upper panels in (b) and (c) display time-of-flight images of BECs oscillating in the $\text{X}$ and $\text{Y}$ modes, respectively, at different hold times. The lower panels show the time evolution of the normalized condensate widths in the $x$ and $y$ directions, respectively.
Data points represent single measurements and the solid lines indicate damped sinusoidal fits to the experimental data.
} 
    \label{quadrupole mode}
\end{figure} 

To derive the low-lying collective excitation modes of the BEC, we consider a small variation of the number density from its equilibrium value, which oscillates in time with the angular frequency $\omega$. By replacing the number density with $n=n_0(\bold{r})+\delta n(\boldsymbol{r})\cos{\omega t}$ in Eqs.~(\ref{conti}) and (\ref{hydro}), we obtain, to the linear order of $\delta n$, 
\begin{equation}
\omega^2\delta n=\frac{1}{m}\left[ \boldsymbol{\nabla} V_\text{ext} \cdot \boldsymbol{\nabla} \delta n-\left(\mu-V_\text{ext}\right)\boldsymbol{\nabla}^2\delta n \right],
\label{deter}
\end{equation}
and the ansatz $\delta n(\boldsymbol{r})=b_0+\sum_{\alpha} b_\alpha r_\alpha^2 $ for quadrupole excitations leads to the characteristic equations 
\begin{align}
    (2\omega_\alpha^2 - \omega^2) b_\alpha +\omega_\alpha^2 \sum_{\beta=x,y,z}  b_\beta &= 0,
    \label{determinant}
     \\
    \omega^2 b_0 + \frac{2\mu}{m} \sum_{\alpha} b_\alpha &=0.
\end{align}
Eq.~(\ref{determinant}) presents three independent sets of solutions for $(\omega, b_\alpha)$, corresponding to different modes of quadrupole excitation. 
From the linearized form of Eq.~(\ref{conti}) for small values of $\delta n$ and $\boldsymbol{v}$, the velocity field in the oscillating BEC is given by 
\begin{equation}
    \label{velocity}
    \boldsymbol{v}(\boldsymbol{r},t) = - \frac{2g}{m\omega} (b_x x, b_y y, b_z z)\sin{\omega t}.
\end{equation}

The three quadrupole excitation modes for an oblate trapping potential with $\omega_x < \omega_{y}<\omega_z$ are depicted in Fig.~\ref{quadrupole mode}(a), which we refer to as the $\text{X, Y}$, and $\text{Z}$ modes, respectively, indicating the axis of dominant width oscillations. 
In the $\text{X (Y)}$ mode, the condensate exhibits out-of-phase (in-phase) width oscillations along the $x$ and $y$ directions. 
We experimentally identified the two quadrupole oscillation modes by applying trap modulations at various frequencies and measuring the width oscillations of the condensate along the $x$ and $y$ directions [Figs.~\ref{quadrupole mode}(b) and \ref{quadrupole mode}(c)]. 
For our samples at the lowest temperature $\zeta_{th}<0.1$ and $\mathcal{D}_{s}=1$, 
the oscillation frequencies were measured and found to be $\omega/2\pi= 7.0(1)$~Hz and $15.0(2)$~Hz for the $\text{X}$ and $\text{Y}$ modes, respectively, which are in good agreement with the predicted values of $7.0$~Hz and $14.7$~Hz from Eq.~(\ref{determinant}) with the trapping frequencies of $(\omega_x,\omega_y,\omega_z)/2\pi= (4.4, 8.9, 415)$~Hz. 

\subsection{Resonance behavior of quadrupole oscillations}
\label{Resonance}

\begin{figure}[b]
\includegraphics[width=9.05cm]{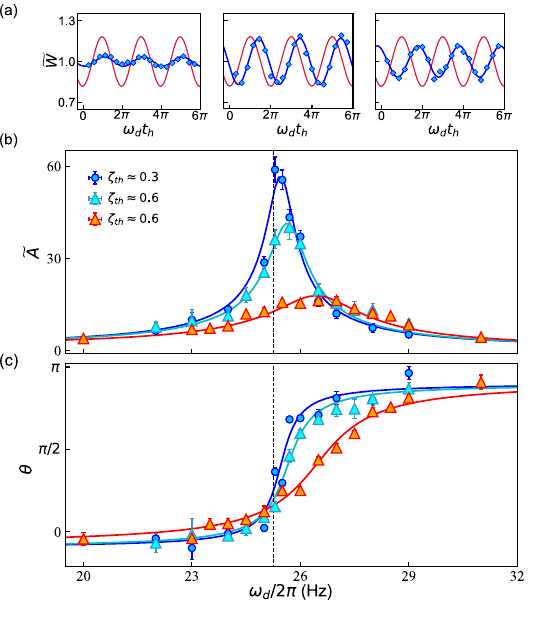}
\centering
\caption{
(a) Time evolution of the normalized condensate width $\widetilde{W}$ (blue diamond) during forced driving for different driving frequencies $\omega_d/2\pi=22$~Hz, 25.5~Hz and 26~Hz from left. The sample was an ordinary BEC sample with thermal fraction $\zeta_{th}= 0.3$. Each data point represents a single measurement. Blue solid lines are sinusoidal functions fitted to the data. Red lines denote guides for oscillations in phase with the trap modulations. 
Responses of BECs to the periodic modulations of the trapping potential. 
(b) Amplitude magnification factor $\widetilde{A}$ and (c) relative phases $\theta$ of the shape oscillations of the driven BECs as functions of the driving frequency $\omega_{d}$ for three different samples: ordinary single-component samples with thermal fractions $\zeta_{th}= 0.3$ (blue) and $= 0.6$ (cyan), and a turbulent BEC sample with $\zeta_{th}= 0.6$ (orange).
Error bars indicate 1$\sigma$ uncertainties including fitting error in (a).
Solid lines show Lorentzian curves in (b) and arctangent functions with offsets in (c), fitted to the corresponding data sets. 
The dashed vertical lines indicate the resonant frequency of the quadrupole oscillation, estimated from the trapping frequencies.
}
\label{fig2}
\end{figure}

We investigated the resonance of the Y quadrupole mode with trap modulations of $U(t)=U_{f}[1+\epsilon \sin (\omega_d t)]$ by measuring the steady-state response of the BEC as a function of the modulation frequency $\omega_d$. 
In this case, we applied the trap modulation for a time of $t_0=50 \frac{2\pi}{\omega_d}$, ensuring that the BEC oscillations reach a quasi-steady state, and continued the modulation in the subsequent hold time $t_h$. We tracked the normalized width $\widetilde{W}(t_h)$ of the BEC along the $y$ direction by taking a time-of-flight image of the sample for various modulation times $t_h$ [Fig.~\ref{fig2}(a)], and determined the in-trap relative amplitude $B$ and phase $\theta$ of the BEC oscillations (see Section~\ref{widthosc}). 

The resonance behavior of the driven BEC was characterized with the amplitude magnification factor $\widetilde{A}(\omega_d)$, which is the ratio of the oscillation amplitude $B$ to the equilibrium displacement for the applied forcing.
Since the change of $W_\text{eq} (\propto U_f^{-1/5})$ for small $\Delta U=\epsilon U_f$ is $\Delta W_{\text{eq}}=-\frac{1}{5}\epsilon W_{\text{eq}}$, the factor is estimated as $\widetilde{A}=5B/\epsilon$.
To ensure our measurements were in the linear response regime, we kept the oscillation amplitude $B$ below 5\%.
We verified that the oscillation amplitude $B$ increases linearly with $\epsilon$ at a modulation frequency of $\omega_d/2\pi=27$~Hz in this limit. 

Figure~\ref{fig2} shows the response spectra $\widetilde{A}(\omega_d)$ and $\theta(\omega_d)$ for three different samples, including non-turbulent BECs prepared without RF spin driving, with respective thermal fractions of $\zeta_{th}=0.32(1)$ (blue markers) and 0.59(1) (cyan), and a turbulent BEC with $\zeta_{th}=0.57(1)$ (orange). The trapping frequencies were $(\omega_x, \omega_y, \omega_z) = 2\pi \times (7.5, 15.3, 730)$~Hz, giving the quadrupole mode frequency $\omega_{\nu,0}/2\pi=25.3$~Hz at zero temperature.
The resonance behavior is evident in $\widetilde{A}$, accompanied by a phase change of $\pi$ in $\theta$ as $\omega_d$ increases.
The resonance frequency $\omega_\nu$ and the spectral width $\Gamma_s$ are determined from the simultaneous fit of $\widetilde{A}(\omega_d)= \frac{A' \omega_\nu^2}{\sqrt{(\omega_d^2-\omega_{\nu}^2)^2+4(\omega_d\Gamma_s)^2}}$ and $\theta(\omega_d)=\frac{\pi}{2}+\tan^{-1}(\frac{\omega_d^2-\omega_{\nu}^2}{2\Gamma_s\omega_d})-\theta_0$ to the experimental data with $A'$ and $\theta_0$ being free parameters. In the fitting, $A' \approx 1.7$ and $\theta_0 \approx 0.1\pi$. The small non-zero value of $\theta_0$ is attributed to a systematic effect in the time-of-flight measurement.
For BECs without turbulence, we obtain $\Gamma_s/2\pi = 0.40(3)$~Hz and $\Delta \omega_\nu/2\pi =(\omega_\nu-\omega_{\nu,0})/2\pi = 0.21(3)$~Hz for a low $\zeta_{th}$, and $\Gamma_s/2\pi = 0.53(5)$~Hz and $\Delta \omega_\nu/2\pi= 0.44(1)$~Hz for a high $\zeta_{th}$.
As $\zeta_{th}$ increases, the resonance frequency shifts upward and the damping increases.
The turbulent BEC demonstrates a more pronounced spectral broadening and a larger shift with $\Gamma_s/2\pi = 1.1(1)$~Hz and $\Delta \omega_\nu/2\pi = 1.6(3)$~Hz.  

We briefly comment on the frequency shifts observed in the driven response spectra. Anomalous frequency shifts and damping of collective modes were first reported in Ref.~\cite{sjin1997temperature}, and subsequent studies identified mean-field coupling between the condensate and thermal cloud as the primary mechanism~\cite{sjackson2002quadrupole,smorgan2003quantitative}. In particular, the direction of the frequency shift can reverse depending on their relative motion~\cite{sjackson2002quadrupole}.
In our driven experiments, the observed blueshift is probably due to this coupling, indicating that a simple one-way Landau damping picture is insufficient. A full explanation would require modeling the long-term dynamical interaction between the condensate and the thermal cloud, such as within the Zaremba-Nikuni-Griffin (ZNG) formalism~\cite{sjackson2002quadrupole}, which is beyond the scope of this work.
We note that in our free-decay experiments, where a short pulse drive is applied at an optimized frequency, no significant temperature-dependent frequency shift was observed.

\begin{figure}[b]
\includegraphics[width=9.0cm]{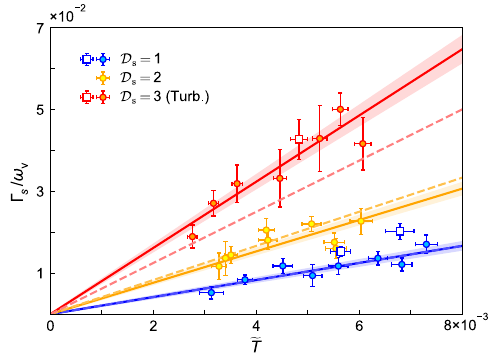}
\centering
\caption{Relative spectral width $\Gamma_s/\omega_\nu$ (open markers) as a function of reduced temperature $\widetilde{T}$, extracted from the response spectra shown in Fig.~\ref{fig2}. For comparison, the solid circles represents the relative damping rates $\Gamma/\omega_\nu$ from free-decay measurements, as also shown in Fig.~3(b). 
The solid and dashed lines are the same as those described in the caption of Fig.~3.
}
\label{figs3}
\end{figure}

In Fig.~\ref{figs3}, we present the measured values of $\Gamma_s/\omega_\nu$, together with the relative damping rates $\Gamma/\omega_\nu$ from Fig.~3(b). The slightly higher values of $\Gamma_s/\omega_\nu$ are probably due to the steady-state relative motion between the condensate and the thermal cloud during the driven measurements.

\subsection{Quantum Reynolds-Averaged Navier-Stokes equation}
\label{QRANS}

Reynolds-averaged Navier-Stokes (RANS) equations are a widely used approach for modeling turbulent flows~\cite{spope2000turbulent}. They are derived by decomposing the instantaneous quantities, such as the velocity and pressure, into mean and fluctuating components, allowing the effects of turbulence to be captured in a time-averaged manner.

Let us consider the quantum Navier-Stokes equation in Eq.~(\ref{hydro}) and a situation in which the velocity $\boldsymbol{v}$ can be decomposed into the mean velocity $\overline{\boldsymbol{v}}$ and the fluctuating part $\boldsymbol{v'}$ under the assumption of isotropic and homogeneous turbulence. 
Taking the time-averaged form of the equation, where the first-order fluctuation terms cancel out, we obtain the RANS equation as
\begin{equation}
\frac{\partial \overline{v}_\alpha}{\partial t}+(\overline{\boldsymbol{v}}\cdot\boldsymbol{\nabla})\overline{v}_\alpha=-\frac{1}{m}\frac{\partial p}{\partial r_\alpha}-\frac{1}{\overline{n}}\sum_{\beta}\frac{\partial}{\partial r_\beta}\overline{n 
 v'_\alpha v'_\beta},
\label{RANS}
\end{equation}
where $p=V_\text{ext}+g\overline{n}$ is the effective pressure, neglecting the quantum pressure.
The upper bar indicates the mean over time averaging.
The last term, $\overline{n v'_\alpha v'_\beta}$ , is known as the Reynolds stress tensor and its value cannot be determined a priori, a situation commonly termed the turbulence closure problem. 
In an analogy to the molecular viscosity that arises from molecular motion in a gas, the Boussinesq hypothesis was proposed~\cite{sschmitt2007about}, stating that Reynolds stresses can be modeled as proportional to the mean strain rate and thus introduce turbulent viscosity in the mean flow motion. 
According to this model, the Reynolds stress tensor is approximated as 
\begin{equation}
\label{Reynolds viscosity}
-\overline{nv'_\alpha v'_\beta}=\overline{n} \nu_\text{T}\sigma_{\alpha\beta}-\frac{2}{3}k \delta_{\alpha\beta},
\end{equation}
where 
$\nu_\text{T}$ is the turbulent viscosity, $\sigma_{\alpha\beta}=\frac{\partial \overline{v}_\alpha}{\partial r_\beta}+\frac{\partial \overline{v}_\beta}{\partial r_\alpha}-\frac{2}{3} \delta_{\alpha \beta}\boldsymbol{\nabla}\cdot\overline{\boldsymbol{v}}$ is the traceless mean strain rate tensor, and $k=\frac{1}{2}  \sum_\alpha \overline{nv'^2_\alpha}$ is the turbulent kinetic energy.

\subsection{Turbulent damping of collective oscillations}

Integrating the turbulent viscosity from Eq.~(\ref{Reynolds viscosity}) into the RANS equation, the linearized equation of motion for a small variation of the mean number density, $\delta \overline{n}$, remains identical to Eq.~(\ref{deter}). 
This means that, despite the perturbation of the turbulent viscosity, the quadrupole excitation modes for a trapped BEC continue to hold for the averaged number density and velocity field, and their oscillation frequency and profile $(\omega,b_\alpha)$ remain consistent with those derived in Section~\ref{Quadrupole mode calc}. 

The primary effect of turbulent viscosity is the dissipation of energy from the quadrupole oscillations.
The total mechanical energy $E$ for the collective oscillations with respect to the equilibrium state is twice the time average of the kinetic energy $E_{kin}(t) = \int d\boldsymbol{r} \frac{1}{2}m \overline{n}(\boldsymbol{r},t)\overline{v}^2(\boldsymbol{r},t)$~\cite{skavoulakis1998} and up to the linear order of $\delta \overline{n}$ and $\overline{\boldsymbol{v}}$, it is given by 
\begin{equation}
\label{energy}
    E=\int d\boldsymbol{r} \left[\frac{1}{2}m n_0 (\boldsymbol{r}) \sum_{\alpha} \left(\frac{2g}{m\omega}\right)^2 b_\alpha^2 r_\alpha^2\right]. 
\end{equation}
The rate of energy dissipation due to viscosity is given by 
\begin{equation}
\label{changerate}
\dot{E}(t) = \int d\boldsymbol{r} \sum_{\alpha, \beta} m{\overline{v}_\alpha} \frac{\partial}{\partial r_\beta}\left(\overline{n} \nu_\text{T}\sigma_{\alpha \beta}-\frac{2}{3}k\delta_{\alpha \beta}\right). 
\end{equation}
To calculate the damping rate of our quadrupole mode, assuming that the amplitude of the quadrupole oscillation does not significantly changed over an oscillation period, the mean energy dissipation rate is expressed as
\begin{equation}
    \label{energyrate}
    \langle \dot{E} \rangle_T=- \int d\boldsymbol{r} \left[\frac{1}{2} m n_0(\boldsymbol{r}) \nu_\text{T} \sum_{\alpha,\beta=1}^3 \left\langle\sigma_{\alpha\beta}^2(\boldsymbol{r},t)\right\rangle_T\right],
\end{equation}
where $\langle\ldots\rangle_T$ denotes the time average over a period of one oscillation~\cite{slandau2013fluid}. 

\begin{figure}[t]
    \includegraphics[width=8.6cm]{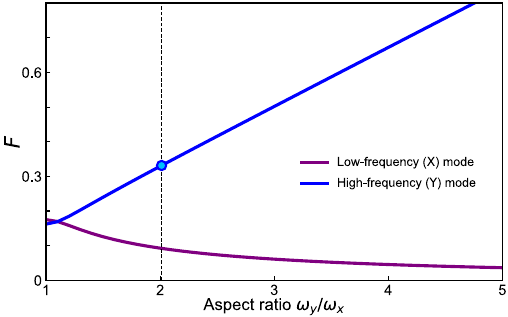}
    \centering
\caption{Characteristic strain values $F(\omega_\alpha, b_\alpha)$ for the $\text{X}$ (low-frequency) and $\text{Y}$ (high-frequency) modes are plotted as functions of the trap aspect ratio $\lambda = \omega_y/\omega_x$. 
The parameter $\omega_z$ and $\sqrt{\omega_x \omega_y}$ are kept constant in the experiment. 
The dashed vertical line indicates the experimental trap condition, where the $\text{Y}$ mode (blue circle) exhibits stronger shear stress than the $\text{X}$ mode. 
}
\label{fig7}
\end{figure}

The amplitude decay rate $\Gamma_\text{T}$ of the collective oscillations is estimated as $\Gamma_\text{T}=-\frac{1}{2}\langle\Dot{E}\rangle_T / E$ and for the velocity field $\overline{\boldsymbol{v}}$ of the quadrupole mode in Eq.~(\ref{velocity}), the damping rate is given by 
\begin{equation}
\label{turbulencedamping}
    \Gamma_\text{T} = \frac{7\nu_T}{3\Bar{R}^2}\frac{[3\sum_{\alpha} b_\alpha^2-(\sum_\alpha b_\alpha)^2]}{\sum_\alpha b_\alpha^2(\Bar{\omega}/\omega_\alpha)^2} 
    = \frac{7 \nu_\text{T}}{3\Bar{R}^2} F(\omega_\alpha, b_\alpha),
\end{equation}
with $\Bar{R} = (R_x R_y R_z)^{1/3}$ and $\Bar{\omega} = (\omega_x \omega_y \omega_z)^{1/3}$, where $R_\alpha$ denotes the TF radius along the $\alpha$ axis.  
$F(\omega_\alpha, b_\alpha)$ is a dimensionless parameter that is determined by the trapping frequencies and the collective oscillation profile, representing the degree of shear strain in the given collective excitation mode.

In Fig.~\ref{fig7}, we plot the values $F$ for the $\text{X}$ and $\text{Y}$ quadrupole modes as functions of the aspect ratio of the transverse trapping frequencies $\lambda=\omega_y/\omega_x$.
In general, the $\text{Y}$ mode, which exhibits in-phase oscillation on the $xy$ plane, displays stronger shear stress than the $\text{X}$ mode with out-of-phase oscillations, due to the motion along the $z$ direction [Fig.~\ref{quadrupole mode}(a)]. For the trap parameters used in our experiment, we find $F=0.345$ for the Y mode and $F=0.096$ for the X mode, implying that $\Gamma_\text{T}$ for the Y mode is approximately three times greater than that of the X mode for a given $\nu_\text{T}$.

For comparison, we also measured the damping rate of the X mode. The experimental protocol was identical to that used for the Y mode, except that the drive frequency $\omega_d$ was tuned specifically for the X mode, and the condensate width $\widetilde{W}$ was measured along the $x$ direction, where the X mode exhibits a larger oscillation amplitude [Fig.~\ref{quadrupole mode}(b)]. The trap frequencies were $(\omega_x,\omega_y,\omega_z) = 2\pi \times (7.5, 15.3, 730)$ Hz. For $\mathcal{D}_{s}=3$ turbulent samples at a reduced temperature of $\widetilde{T}=4.6(1)\times10^{-3}$, we measured  $\Gamma/2\pi=0.48(11)$~Hz and $\omega_\nu/2\pi=11.7(1)$~Hz. For $\mathcal{D}_{s}=1$ samples at $\widetilde{T}=6.0(2)\times10^{-3}$, the corresponding values were $\Gamma/2\pi=0.21(7)$~Hz and $\omega_\nu/2\pi=11.6(1)$~Hz. 

Using the relation of $A_\nu^{(0)}=\Gamma/(\omega_\nu \widetilde{T})$ for the $\mathcal{D}_{s}=1$ case, the excess damping in the $\mathcal{D}_s=3$ turbulent sample is determined from $\Gamma_\text{T}=\Gamma-3 A_\nu^{(0)} \widetilde{T} \omega_\nu$ [Eq.~(2)] as described in the main text. From the two measurements, we extract the damping rate of $\Gamma_\text{T}=-0.01\pm0.19$ Hz. Given that $\Gamma_\text{T}/2\pi\approx 0.2$~Hz was observed for the Y mode, we expect $\Gamma_\text{T}/2\pi\approx 0.06$~Hz for the X mode, which lies below the resolution of our current experiment. A more systematic study exploring the dependence of $\Gamma_\text{T}$ on trap geometry and shear stress would be needed to rigorously validate the model.


\begin{thebibliography}{99}

\bibitem{pope2000turbulent}
S. B. Pope,
\textit{Turbulent Flows},
(Cambridge University Press, Cambridge, 2000).

\bibitem{schmitt2007about}
F. G. Schmitt,
About Boussinesq's turbulent viscosity hypothesis: historical remarks and a direct evaluation of its validity,
C. R. Méc. {\bf 335}, 617 (2007).

\bibitem{leschziner2002turbulence}
M. A. Leschziner and D. Drikakis,
Turbulence modelling and turbulent-flow computation in aeronautics,
Aeronaut. J. {\bf 106}, 349 (2002).

\bibitem{bhide2021improved}
K. Bhide, K. Siddappaji, S. Abdallah, and K. Roberts,
Improved supersonic turbulent flow characteristics using non-linear eddy viscosity relation in RANS and HPC-enabled LES,
Aerospace {\bf 8}, 352 (2021).

\bibitem{chen1995comparison}
Q. Chen,
Comparison of different k-$\varepsilon$ models for indoor air flow computations,
Numer. Heat Transf. B: Fundam. {\bf 28}, 353 (1995)


\bibitem{vinen2002quantum}
W. F. Vinen and J. J. Niemela,
Quantum turbulence,
J. Low Temp. Phys. {\bf 128}, 167 (2002).

\bibitem{barenghi2014experimental}
C. F. Barenghi, V. S. L'vov, and P.-E. Roche,
Experimental, numerical, and analytical velocity spectra in turbulent quantum fluid,
Proc. Natl. Acad. Sci. USA {\bf 111}, 4683(2014).

\bibitem{pu1999coherent}
H. Pu, C. K. Law, J. H. Eberly, and N. P. Bigelow,
Coherent disintegration and stability of vortices in trapped Bose condensates,
Phys. Rev. A {\bf 59}, 1533 (1999).



\bibitem{bewley2008characterization}
G. P. Bewley, M. S. Paoletti, K. R. Sreenivasan, and D. P. Lathrop,
Characterization of reconnecting vortices in superfluid helium,
Proc. Natl. Acad. Sci. USA {\bf 105}, 13707 (2008).

\bibitem{hall1956rotation}
H. E. Hall and W. F. Vinen,
The rotation of liquid helium II. II. The theory of mutual friction in uniformly rotating helium II,
Proc. R. Soc. Lond. A {\bf 238}, 215 (1956).

\bibitem{skrbek2021phenomenology}
L. Skrbek, D. Schmoranzer, \v{S}. Midlik, and K. R. Sreenivasan,
Phenomenology of quantum turbulence in superfluid helium,
Proc. Natl. Acad. Sci. USA {\bf 118}, e2018406118 (2021).






\bibitem{maurer1998local}
J. Maurer and P. Tabeling,
Local investigation of superfluid turbulence,
EPL {\bf 43}, 29 (1998).


\bibitem{navon2016emergence}
N. Navon, A. L. Gaunt, R. P. Smith, and Z. Hadzibabic,
Emergence of a turbulent cascade in a quantum gas,
Nature {\bf 539}, 72 (2016).

\bibitem{zhao2025kolmogorov}
M. Zhao, J. Tao, and I. B. Spielman,
Kolmogorov scaling in turbulent 2D Bose-Einstein condensates,
Phys. Rev. Lett. {\bf 134}, 083402 (2025).

\bibitem{brewer1961heat}
D. F. Brewer and D. O. Edwards,
Heat conduction by liquid helium II in capillary tubes. I: Transition to supercritical conduction,
Philos. Mag. {\bf 6}, 775 (1961).

\bibitem{eddyviscosityHe4}
R. K. Childers and J. T. Tough,
Eddy viscosity of turbulent superfluid $^{4}\mathrm{He}$,
Phys. Rev. Lett. {\bf 35}, 527 (1975).

\bibitem{kashiwamura1999effects}
S. Kashiwamura, K. Miyake, K. Yamada, and M. Yamaguchi,
Effects of eddy viscosity of turbulent superfluid in capillary flows,
Physica B {\bf 270}, 60 (1999).

\bibitem{mossman2018dissipative}
M. E. Mossman, M. A. Hoefer, K. Julien, P. G. Kevrekidis, and P. Engels,
Dissipative shock waves generated by a quantum-mechanical piston,
Nat. Commun. \textbf{9}, 4665 (2018).

\bibitem{andersson2007superfluid}
N. Andersson, T. Sidery, and G. L. Comer,
Superfluid neutron star turbulence,
Mon. Not. R. Astron. Soc. {\bf 381}, 747 (2007).

\bibitem{melatos2014pulsar}
A. Melatos and B. Link,
Pulsar timing noise from superfluid turbulence,
Mon. Not. R. Astron. Soc. {\bf 437}, 21 (2014).


\bibitem{fujimoto2014}
K. Fujimoto and M. Tsubota,
Spin-superflow turbulence in spin-1 ferromagnetic spinor Bose-Einstein condensates,
Phys. Rev. A {\bf 90}, 013629 (2014).

\bibitem{hong2023spin}
D. Hong, J. Lee, J. Kim, J. H. Jung, K. Lee, S. Kang, and Y. Shin,
Spin-driven stationary turbulence in spinor Bose-Einstein condensates,
Phys. Rev. A {\bf 108}, 013318 (2023).

\bibitem{jung2023random}
J. H. Jung, J. Lee, J. Kim, and Y. Shin,
Random spin textures in turbulent spinor Bose-Einstein condensates,
Phys. Rev. A {\bf 108}, 043309 (2023).




\bibitem{fedichev1998damping}
P. O. Fedichev, G. V. Shlyapnikov, and J. T. M. Walraven,
Damping of low-energy excitations of a trapped Bose-Einstein condensate at finite temperatures,
Phys. Rev. Lett. {\bf 80}, 2269 (1998).

\bibitem{pethick2002bose}
C. J. Pethick and H. Smith,
\textit{Bose-Einstein Condensation in Dilute Gases},
(Cambridge University Press, Cambridge, 2002).



\bibitem{landau1987fluid}
L. D. Landau and E. M. Lifshitz,
\textit{Fluid Mechanics: Landau and Lifshitz: Course of Theoretical Physics, Volume 6},
(Elsevier Science, Amsterdam, 1987).





\bibitem{rautenberg2020classical}
M. Rautenberg and M. G\"arttner,
Classical and quantum chaos in a three-mode bosonic system,
Phys. Rev. A {\bf 101}, 053604 (2020).

\bibitem{evrard2021many}
B. Evrard, A. Qu, J. Dalibard, and F. Gerbier,
From many-body oscillations to thermalization in an isolated spinor gas,
Phys. Rev. Lett. {\bf 126}, 063401 (2021).

\bibitem{kim2024chaos}
J. Kim, J. Jung, J. Lee, D. Hong, and Y. Shin,
Chaos-assisted turbulence in spinor Bose-Einstein condensates,
Phys. Rev. Res. {\bf 6}, L032030 (2024).


\bibitem{landaudamping}
L. D. Landau,
On the vibrations of the electronic plasma,
J. Phys. USSR {\bf 10}, 25 (1946).

\bibitem{chevy2002breathing}
F. Chevy, V. Bretin, P. Rosenbusch, K. W. Madison, and J. Dalibard,
Transverse Breathing Mode of an Elongated Bose-Einstein Condensate.
Phys. Rev. Lett. {\bf 88}, 250402 (2002).

\bibitem{jackson2003landau}
B. Jackson and E. Zaremba,
Landau damping in trapped Bose condensed gases,
New J. Phys. {\bf 5}, 88 (2003).

\bibitem{jin1997temperature}
D. S. Jin, M. R. Matthews, J. R. Ensher, C. E. Wieman, and E. A. Cornell,
Temperature-dependent damping and frequency shifts in collective excitations of a dilute Bose-Einstein condensate,
Phys. Rev. Lett. {\bf 78}, 764 (1997).

\bibitem{stamper1998collisionless}
D. M. Stamper-Kurn, H.-J. Miesner, S. Inouye, M. R. Andrews, and W. Ketterle,
Collisionless and hydrodynamic excitations of a Bose-Einstein condensate,
Phys. Rev. Lett. {\bf 81}, 500 (1998).

\bibitem{jackson2002quadrupole}
B. Jackson and E. Zaremba,
Quadrupole collective modes in trapped finite-temperature Bose-Einstein condensates,
Phys. Rev. Lett. {\bf 88}, 180402 (2002).

\bibitem{morgan2003quantitative}
S. A. Morgan, M. Rusch, D. A. W. Hutchinson, and K. Burnett,
Quantitative test of thermal field theory for Bose-Einstein condensates,
Phys. Rev. Lett. {\bf 91}, 250403 (2003).

\bibitem{meppelink2009damping}
R. Meppelink, S. B. Koller, J. M. Vogels, H. T. C. Stoof, and P. van der Straten,
Damping of superfluid flow by a thermal cloud,
Phys. Rev. Lett. {\bf 103}, 265301 (2009).

\bibitem{yuen2015enhanced}
B. Yuen, I. J. M. Barr, J. P. Cotter, E. Butler, and E. A. Hinds,
Enhanced oscillation lifetime of a Bose--Einstein condensate in the 3D/1D crossover,
New J. Phys. {\bf 17}, 093041 (2015).

\bibitem{SM}
See Supplemental Material for details of the experimental
methods of sample preparation and imaging, a description of the quadrupole modes of highly oblate condensates, experimental data for forced oscillations, and the derivation of $\nu_\text{T}$ using the RANS equation.


\bibitem{stringari1996collective}
S. Stringari,
Collective excitations of a trapped Bose-condensed gas,
Phys. Rev. Lett. {\bf 77}, 2360 (1996).



\bibitem{Heiselberg2004collective}
H. Heiselberg,
Collective modes of trapped gases at the BEC-BCS crossover,
Phys. Rev. Lett. {\bf 93}, 040402 (2004).

\bibitem{cao2011}
C. Cao, E. Elliott, J. Joseph, H. Wu, J. Petricka, T. Sch\"afer, and J. E. Thomas,
Universal quantum viscosity in a unitary Fermi gas,
Science {\bf 331}, 58 (2011).

\bibitem{vogt2012scale}
E. Vogt, M. Feld, B. Fr\"ohlich, D. Pertot, M. Koschorreck, and M. K\"ohl,
Scale invariance and viscosity of a two-dimensional Fermi gas,
Phys. Rev. Lett. {\bf 108}, 070404 (2012).




\bibitem{kawaguchi2012spinor}
Y. Kawaguchi and M. Ueda,
Spinor Bose--Einstein condensates,
Phys. Rep. {\bf 520}, 253 (2012).





\bibitem{bhattacherjee2014damping}
A. B. Bhattacherjee,
Damping in two-component Bose gas,
Mod. Phys. Lett. B {\bf 28}, 1450029 (2014).

\bibitem{stenger1998spin}
J. Stenger, S. Inouye, D. M. Stamper-Kurn, H.-J. Miesner, A. P. Chikkatur, and W. Ketterle,
Spin domains in ground-state Bose-Einstein condensates,
Nature {\bf 396}, 345 (1998).

\bibitem{gerbier2006resonant}
F. Gerbier, A. Widera, S. Fölling, O. Mandel, and I. Bloch,
Resonant control of spin dynamics in ultracold quantum gases by microwave dressing,
Phys. Rev. A {\bf 73}, 041602 (2006).

\bibitem{seo2015half}
S. W. Seo, S. Kang, W. J. Kwon, and Y. Shin,
Half-quantum vortices in an antiferromagnetic spinor Bose-Einstein condensate,
Phys. Rev. Lett. {\bf 115}, 015301 (2015).


\bibitem{duine2009spin}
R. A. Duine and H. T. C. Stoof,
Spin drag in noncondensed Bose gases,
Phys. Rev. Lett. {\bf 103}, 170401 (2009).

\bibitem{jpsj2013}
M. Sadgrove, Y. Eto, S. Sekine, H. Suzuki, and T. Hirano,
Ramsey interferometry using the Zeeman sublevels in a spin-2 Bose gas,
J.~Phys.~Soc.~Jpn. {\bf 82}, 094002 (2013).

\bibitem{knoop2011}
S. Knoop, T. Schuster, R. Scelle, A. Trautmann, J. Appmeier, M. K. Oberthaler, E. Tiesinga and E. Tiemann,
Feshbach spectroscopy and analysis of the interaction potentials of ultracold sodium,
Phys. Rev. A {\bf 83}, 042704 (2011).

\bibitem{footnote2}
In the Thomas-Fermi limit, 
$\mu = \frac{\hbar \overline{\omega}}{2} (15 N_c a/\overline{a})^{2/5}$ 
and 
$n_0=\mu \frac{m}{4\pi \hbar^2 a}$ 
with $\overline{a}=\Bigl(\frac{\hbar}{m\overline{\omega}}\Bigr)^{1/2}$.

\bibitem{footnote_scat}
Because $a$ for $m_F=-1$ is largest, adopting it would underestimate $A_\nu$ for $\mathcal{D}_{s}=3$.

\bibitem{footnote_SG}
Because of magnetic-field curvature, the $m_F=\pm1$ components were stretched slightly differently during spin separation.






\bibitem{footnote3}
In our experiment, $\mu/(\hbar \omega_z) \sim 2$ and $k_\text{B}T/(\hbar \omega_z)$ varied between 7 and 13.



\bibitem{kavoulakis1998}
G. M. Kavoulakis, C. J. Pethick, and H. Smith,
Damping of hydrodynamic modes in a trapped Bose gas above the Bose-Einstein transition temperature,
Phys. Rev. A {\bf 57}, 2938 (1998).

\bibitem{walmsley2008quantum}
P. M. Walmsley and A. I. Golov,
Quantum and quasiclassical types of superfluid turbulence,
Phys. Rev. Lett. {\bf 100}, 245301 (2008).

\bibitem{babuin2014effective}
S. Babuin, E. Varga, L. Skrbek, E. L\'ev\^eque, and P.-E. Roche,
Effective viscosity in quantum turbulence: A steady-state approach,
Europhys. Lett. {\bf 106}, 24006 (2014).

\bibitem{zmeev2015dissipation}
D. E. Zmeev, P. M. Walmsley, A. I. Golov, P. V. E. McClintock, S. N. Fisher, and W. F. Vinen,
Dissipation of quasiclassical turbulence in superfluid $^{4}$He,
Phys. Rev. Lett. {\bf 115}, 155303 (2015).

\bibitem{shukla2019quantitative}
V. Shukla, P. D. Mininni, G. Krstulovic, P. C. Di Leoni, and M. E. Brachet,
Quantitative estimation of effective viscosity in quantum turbulence,
Phys. Rev. A {\bf 99}, 043605 (2019).

\bibitem{amette2025vortex}
J. Amette Estrada, M. E. Brachet, and P. D. Mininni,
Vortex-lattice melting and critical temperature shift in rotating Bose-Einstein condensates,
Phys. Rev. A {\bf 111}, 023304 (2025).

\bibitem{shats2010turbulence}
M. Shats, D. Byrne, and H. Xia,
Turbulence decay rate as a measure of flow dimensionality,
Phys. Rev. Lett. {\bf 105}, 264501 (2010).

\bibitem{lobser2015observation}
D. S. Lobser, A. E. S. Barentine, E. A. Cornell, and H. J. Lewandowski,
Observation of a persistent non-equilibrium state in cold atoms,
Nat. Phys. {\bf 11}, 1009 (2015).

\end{thebibliography}

\begin{thebibliography}{99}



\bibitem{sgerbier2006resonant}
F. Gerbier, A. Widera, S. Fölling, O. Mandel, and I. Bloch,
Resonant control of spin dynamics in ultracold quantum gases by microwave dressing,
Phys. Rev. A {\bf 73}, 041602 (2006).

\bibitem{sseo2015half}
S. W. Seo, S. Kang, W. J. Kwon, and Y. Shin,
Half-quantum vortices in an antiferromagnetic spinor Bose-Einstein condensate,
Phys. Rev. Lett. {\bf 115}, 015301 (2015).

\bibitem{shong2023spin}
D. Hong, J. Lee, J. Kim, J. H. Jung, K. Lee, S. Kang, and Y. Shin,
Spin-driven stationary turbulence in spinor Bose-Einstein condensates,
Phys. Rev. A {\bf 108}, 013318 (2023).

\bibitem{skim2024chaos}
J. Kim, J. Jung, J. Lee, D. Hong, and Y. Shin,
Chaos-assisted turbulence in spinor Bose-Einstein condensates,
Phys. Rev. Res. {\bf 6}, L032030 (2024).

\bibitem{srautenberg2020classical}
M. Rautenberg and M. G\"arttner,
Classical and quantum chaos in a three-mode bosonic system,
Phys. Rev. A {\bf 101}, 053604 (2020).

\bibitem{sevrard2021many}
B. Evrard, A. Qu, J. Dalibard, and F. Gerbier,
From many-body oscillations to thermalization in an isolated spinor gas,
Phys. Rev. Lett. {\bf 126}, 063401 (2021).

\bibitem{sjung2023random}
J. H. Jung, J. Lee, J. Kim, and Y. Shin,
Random spin textures in turbulent spinor Bose-Einstein condensates,
Phys. Rev. A {\bf 108}, 043309 (2023).

\bibitem{skawaguchi2012spinor}
Y. Kawaguchi and M. Ueda,
Spinor Bose--Einstein condensates,
Phys. Rep. {\bf 520}, 253 (2012).

\bibitem{sstringari1996collective}
S. Stringari, 
Collective excitations of a trapped Bose-condensed gas,
Phys. Rev. Lett. {\bf 77}, 2360 (1996).

\bibitem{sHeiselberg2004collective}
H. Heiselberg,
Collective modes of trapped gases at the BEC-BCS crossover,
Phys. Rev. Lett. {\bf 93}, 040402 (2004).

\bibitem{sjin1997temperature}
D. S. Jin, M. R. Matthews, J. R. Ensher, C. E. Wieman, and E. A. Cornell,
Temperature-dependent damping and frequency shifts in collective excitations of a dilute Bose-Einstein condensate,
Phys. Rev. Lett. {\bf 78}, 764 (1997).

\bibitem{sjackson2002quadrupole}
B. Jackson and E. Zaremba,
Quadrupole collective modes in trapped finite-temperature Bose-Einstein condensates,
Phys. Rev. Lett. {\bf 88}, 180402 (2002).

\bibitem{smorgan2003quantitative}
S. A. Morgan, M. Rusch, D. A. W. Hutchinson, and K. Burnett,
Quantitative test of thermal field theory for Bose-Einstein condensates,
Phys. Rev. Lett. {\bf 91}, 250403 (2003).

\bibitem{spope2000turbulent}
S. B. Pope,
\textit{Turbulent Flows},
(Cambridge University Press, Cambridge, 2000).

\bibitem{sschmitt2007about}
F. G. Schmitt,
About Boussinesq's turbulent viscosity hypothesis: historical remarks and a direct evaluation of its validity,
C. R. Méc. {\bf 335}, 617 (2007).

\bibitem{skavoulakis1998}
G. M. Kavoulakis, C. J. Pethick, and H. Smith,
Damping of hydrodynamic modes in a trapped Bose gas above the Bose-Einstein transition temperature,
Phys. Rev. A {\bf 57}, 2938 (1998).

\bibitem{slandau2013fluid}
L. D. Landau and E. M. Lifshitz,
\textit{Fluid Mechanics: Landau and Lifshitz: Course of Theoretical Physics, Volume 6}, 
(Elsevier Science, Amsterdam, 2013).

\end{thebibliography}
\end{document}